\documentclass[aps,amsfonts,reprint,tightenlines,amssymb,superscriptaddress,twocolumn]{revtex4-2}
\usepackage[T1]{fontenc}    
\usepackage{amsfonts} 
\usepackage{amsmath,amsbsy,amssymb,graphicx,enumerate,latexsym,bm,ulem,multirow,listings}                       
\usepackage{times} 
\usepackage{color}

\usepackage[pagebackref=false]{hyperref} % choice of color for hyperref links 
\hypersetup{
    bookmarks=true,         % show bookmarks bar?
    unicode=false,          % non-Latin characters in Acrobat?s bookmarks
    pdftoolbar=true,        % show Acrobat?s toolbar?
    pdfmenubar=true,        % show Acrobat?s menu?
    pdffitwindow=false,     % window fit to page when opened
    pdfstartview={FitH},    % fits the width of the page to the window
    pdfkeywords={keyword1} {key2} {key3}, % list of keywords
    pdfnewwindow=true,      % links in new window
    colorlinks=true,       % false: boxed links; true: colored links
    linkcolor=red,          % color of internal links (change box color with linkbordercolor)
    citecolor=blue,        % color of links to bibliography
    filecolor=magenta,      % color of file links
    urlcolor=cyan,           % color of external links         
}

\makeatletter
\newsavebox{\@brx}
\newcommand{\llangle}[1][]{\savebox{\@brx}{\(\m@th{#1\langle}\)}%
  \mathopen{\copy\@brx\kern-0.5\wd\@brx\usebox{\@brx}}}
\newcommand{\rrangle}[1][]{\savebox{\@brx}{\(\m@th{#1\rangle}\)}%
  \mathclose{\copy\@brx\kern-0.5\wd\@brx\usebox{\@brx}}}
\makeatother
\begin{document}
\title{Double Nuclear Spin Relaxation in Hybrid Quantum Hall Systems}               
\author{M.~H.~Fauzi} \email{moha065@lipi.go.id}
\affiliation{Center for Spintronics Research Network, Tohoku University, Sendai 980-8577, Japan }
\affiliation{Research Center for Physics, Indonesian Institute of Sciences, South Tangerang City, Banten 15314, Indonesia }
\author{William J. Munro}  
\affiliation{National Institute of Informatics, 2-1-2 Hitotsubashi, Chiyoda-ku, Tokyo 101-8430, Japan}
\affiliation{NTT Basic Research Laboratories \& NTT Research Center for Theoretical Quantum Physics, NTT Corporation, 3-1 Morinosato-Wakamiya, Atsugi-shi, Kanagawa, 243-0198, Japan }        
\author{Kae~Nemoto}
\email{nemoto@nii.ac.jp}
\affiliation{National Institute of Informatics, 2-1-2 Hitotsubashi, Chiyoda-ku, Tokyo 101-8430, Japan}
\author{Y.~Hirayama}\email{yoshiro.hirayama.d6@tohoku.ac.jp}
\affiliation{Center for Spintronics Research Network, Tohoku University, Sendai 980-8577, Japan }
\affiliation{Department of Physics, Tohoku University, Sendai 980-8578, Japan}
\affiliation{Center for Science and Innovation in Spintronics (Core Research Cluster), Tohoku University, Sendai 980-8577, Japan}
\date{\today}
\begin{abstract}
Recent advances in quantum engineering have given us the ability to design hybrid systems with novel properties normally not present in the regime they operate in. The coupling of spin ensembles and magnons to microwave resonators has for instance lead to a much richer understanding of collective effects in these systems and their potential quantum applications. We can also hybridize electron and nuclear spin ensembles together in the solid-state regime to investigate collective effects normally only observed in the atomic, molecular and optical world. Here we explore in the solid state regime the dynamics of a double domain nuclear spin ensemble coupled to the Nambu-Goldstone boson  in GaAs semiconductors and show it exhibits both collective and individual relaxation (thermalization) on very different time scales. Further the collective relaxation of the nuclear spin ensemble is what one would expect from superradiant decay. This opens up the possibility for the exploration of novel collective behaviour in solid state systems where the natural energies associated with those spins are much less than the thermal energy.
\end{abstract}
\maketitle
%\section{Introduction}\label{intro}
%\noindent
%\textbf{Recent advances in quantum engineering have given us the ability to design hybrid systems with novel properties normally not present in the regime they operate in. The coupling of spin ensembles and magnons to microwave resonators has for instance lead to a much richer understanding of collective effects in these systems and their potential quantum applications. We can also hybridize electron and nuclear spin ensembles together in the solid-state regime to investigate collective effects normally only observed in the atomic, molecular and optical world. Here we explore in the solid state regime the dynamics of a double domain nuclear spin ensemble coupled to the Nambu-Goldstone boson  in GaAs semiconductors and show it exhibits both collective and individual relaxation (thermalization) on very different time scales. Further the collective relaxation of the nuclear spin ensemble is what one would expect from superradiant decay. This opens up the possibility for the exploration of novel collective behaviour in solid state systems where the natural energies associated with those spins are much less than the thermal energy.} 

It has now be widely accepted that the principles of quantum mechanics will lead to new technologies with capabilities unlike anything seen in our purely classical world \cite{Nielsen200,Spiller2006,Dowling02}. There is the potential to have computational and communication power beyond anything our conventional world can every realize \cite{Gisin2002,Shor1997,Feynman}. The principles also allow us to construct unparalleled quantum sensing and imaging sensitivity \cite{Caves,Giovannetti}. Typically such achievements could be achieved using traditional quantum systems. However  in recent years it has been established that the hybridization of distinct quantum systems has the potential to design composite devices with properties and attributes not normally unavailable in the regime those systems came from  \cite{hybrid1,hybrid2,hybrid3}.  Hybrid quantum systems will exhibit functionalities superior to those in other sub quantum systems (long lived memories for instance) and are  likely to be multitasking \cite{hybrid3}. They will play important roles in engineering  multi-functional quantum devices (quantum transducers for instance) and performing diverse quantum information processing tasks. With the state-of-the-art quantum technology,  hybrid quantum systems are being designed and engineered using many different types of elements ranging from solids to atomic, molecular and optical (AMO) systems \cite{hybrid1,hybrid2,hybrid3,cavityqedreview1,cavityqedreview2,nvcenterreview1,nvcentermechresonatorreview1,strongcoupling1,strongcoupling2,strongcoupling3}. 

Hybrid quantum systems are more however than a tool to create new technologies as they also provide the opportunity to explore quantum many body and non-equilibrium physics in unique regimes or regimes normally not available to those systems \cite{hybrid3,Angerer2017,NVsuperradiantdecay}. We now have the ability to explore quantum phenomena using solid-state systems which have typically been investigated in AMO systems \cite{Angerer2017,Dicke,Gross,solidsuperradiance2016,NVsuperradiantdecay} with the advantage that this  solid-state systems are easier to control, manipulate and measure with high accuracy. Two recent examples include the demonstration of amplitude bistability and superradiance using an electronic spin ensemble coupled to a microwave resonator \cite{Angerer2017,NVsuperradiantdecay}.  The latter case is of particular interest here as it involves the collective behaviour of that ensemble where the superradiant burst of microwave photons occurred ten orders of magnitude faster than the corresponding relaxation of a single nitrogen-vacancy center in diamonds electron spin \cite{NVsuperradiantdecay}. This enabled the dynamics of the process can be investigated. While this was a coherence phenomena, a simple modification using two ensembles instead of one allows it to be a truly collective quantum phenomena with no classical analog \cite{HamaPRL,HamaPRA}. It also raises the question about whether such collective effects can be seen in other solid-state systems which we will explore in this context.  Further by using nuclear spins one would be operating in a regime with significant thermal background, even though the experiment is taking place in a dilution environment \cite{quantumdotreview1,SchuetzetalnuclearspinsinQDNV,GaAsNatmat2013}. 

\begin{figure*}[htb]
\includegraphics[width=0.9\linewidth]{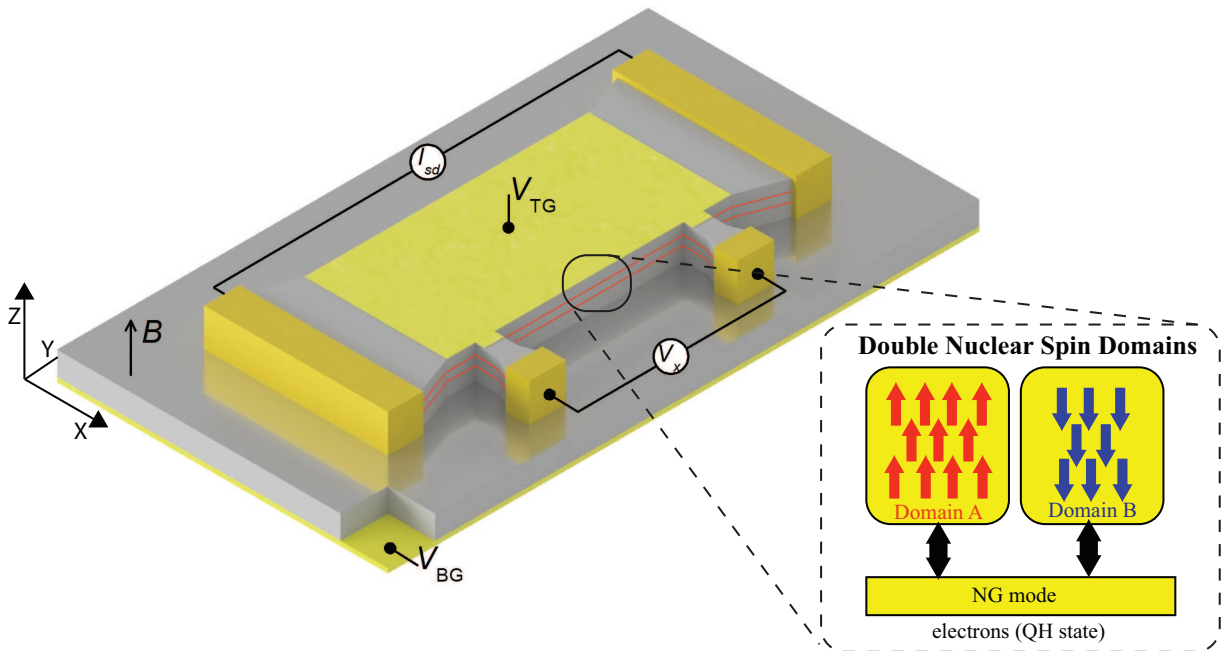}
\caption{Schematic illustration of our hybrid quantum systems device operating at dilution temperatures (50 mK) used to investigate the behaviour of nuclear spins coupled with an electronic reservoir of electrons in QH states through the hyperfine interaction.  Here two identical 20-nm wide GaAs quantum wells, separated by a $2.2$ nm thick insulation layer, are in close proximity to create a strongly coupled bilayer two-dimensional electron gas system. These are indicated by the two red lines in the main illustration. The electronic states in both layers are electrostatically controlled by applying a gate bias between the top gate ($V_{\rm{TG}}$) and bottom gate ($V_{\rm{BG}}$). A static magnetic field $B$ is also applied along the $Z$-axis normal to the device. The inset depicts our  QH system coupled through the hyperfine interaction to the NG boson. The nuclear spins are forming the double-spin domains A and B (two independent ensembles) with the NG boson operating as a reservoir. Domain A is described by the upward red arrows while domain B with the downward blue arrows. The number of spins in each domain is approximately the same.}
\label{Fig01} 
\end{figure*}

The predication and subsequent demonstration of the Quantum Hall (QH) effect establishing macroscopic quantum phenomena in  solid-state two-dimensional electron systems in the presence of  high magnetic field has spawned many important  discoveries in quantum many-body electronic physics  and  topological quantum matter including the quantum Hall ferromagnet and collective excitations such as Nambu-Goldstone (NG) bosons \& skyrmions  \cite{QuantumHall1,QuantumHall2,TISC1,TISC2,Kumada2006}. However the nuclear-electron spin dynamics and its hybridization in the QH system has been rarely explored \cite{GaAsreview}. Coupling of a nuclear spin ensemble to a Nambu-Goldstone boson seems an ideal candidate to explore  collective effects in a solid-state hybrid quantum system \cite{Kumada2006,HamaPRL} and that would be our focus here. We would expect to observe phenomena arising from both collective and individual nuclear spins decoherence and so our initial focus will be on the measured behaviour of this hybrid system - especially the dynamic ones. A Nambu-Goldstone mode is known to cause a rapid nuclear spin relaxation rate \cite{Hashimoto2002,Kumada2006,Kumada2007}. However, the way that the nuclear spin dynamics was measured in the previous reports missed out a key feature, which we address here namely a sudden reconfiguration of nuclear spin polarization.

\begin{figure}[htb]
\includegraphics[width= 0.65\linewidth]{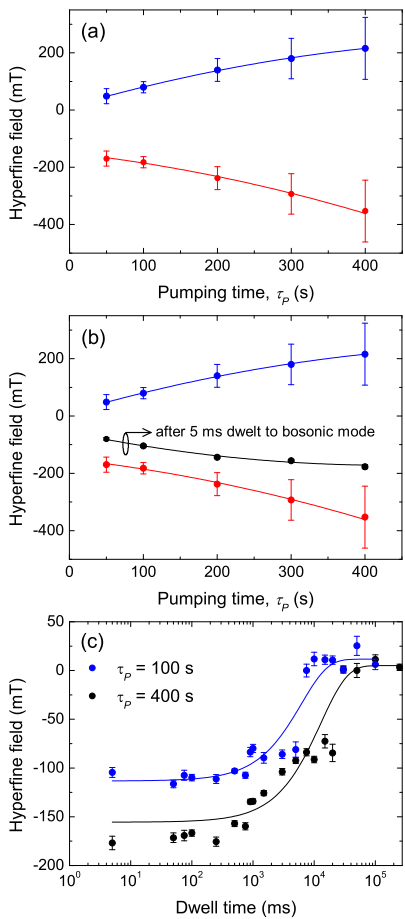}
\caption{Plot  of the initialization of the double-nuclear-spin domain (a) indicating the number of polarized nuclear spins  (measured as a hyperfine field change in mT) versus the time $\tau_{\rm{p}}$ the source-drain current $I_{\text{sd}}$ is applied. At $\tau_{\rm{p}} = 50$s, the number of up nuclear spin polarization $J^z_\uparrow$ (down nuclear spin polarization $J^z_\downarrow$) was $-170$ mT ($+43$ mT) respectively. For $\tau_{\rm{p}} = 400$s, they increased to $-353$ mT ($+215$ mT), a factor of 2 (5) polarization improvement respectively. As a calibration a fully polarized sample would result in a hyperfine field near $5.3$ T \cite{Paget1997} meaning we can estimate the maximum percentage of  spins in the up (down) domain at approximately $6.6\%$ ($4\%$) respectively. In (b) we depict the  total nuclear spin polarization for varying pumping time $\tau_{\rm{p}}$ followed by a $5$ms interaction the NG boson with a subsequent total magnetization measurement (in terms of the hyperfine field strength). For ease of comparison, the initial polarization in panel(a) is replotted together. In (c) we show the nuclear spin relaxation dynamics versus the Dwell time to the NG boson for pumping times $\tau_{\rm{p}}$ $=100$s ($400$s) respectively with the solid lines providing a visual guide. It is clear that at long times the double-nuclear-spin domains reach the same thermal equilibrium. Our minimum Dwell time measurement is $5$ ms restricting our ability to explore the really short time dynamics.}
\label{Fig02} 
\end{figure}

%\section{Experimental Setup and Demonstration}\label{exp}

Let us begin by describing our nuclear spins coupled with electrons in QH states through the Fermi contact hyperfine interaction. Our hybrid quantum system as depicted in Fig. \ref{Fig01} is implemented in a GaAs-based bilayer quantum Hall setting cooled to $50$ mK and threaded with a static magnetic field $B$ applied normal to the device.  This brings the electronic state into the quantum Hall regime where the interplay between charge, spin, and layer degree of freedom creates non-trivial correlated electronic states \cite{Sarma1997,Brey1999} where all nuclear spins can be up or down for instance. For our situation here, the initialization of our double nuclear-spin domains with different spin orientation ($\uparrow$ and $\downarrow$) is achieved by electrical means using a fractional quantum Hall liquid (FQHL) \cite{Jain} developed at the bottom layer, where $2/3$ of the available states in the lowest Landau level are occupied (see also supplementary material). It exhibits two different magnetic phases, spin polarized and unpolarized phases, which are separated by a domain wall at the spin transition point \cite{Stern2004}. Now the injection of a relatively large alternating source-drain current $I_{\text{sd}}$ at the spin transition point creates a bidirectional nuclear spin polarization due to multiple forward scattering events between two neighboring magnetic phases  \cite{FauziAPL, Nick}. As a result the double-nuclear-spin domain is generated with a state of the form $\big{|}\text{DNSD}\big{\rangle} =|\uparrow\ldots\uparrow\rangle_{\text{A}}\otimes|\downarrow\ldots\downarrow\rangle_{\text{B}}$ where the first domain (labelled A) has all spins pointed up $\uparrow$ while the second domain $B$ has all spins pointed down $\downarrow$. We expect the domain to have approximately the same number of nuclear spins in them but a little asymmetry could exist \cite{footnote1}. 

Our  initial state $\big{|}\text{DNSD}\big{\rangle}$ is illustrated in Fig. (\ref{Fig02}a) where we vary the pumping time $\tau_p$ to observe its effect on the magnetization of each domain. Here the blue (red) dots represent the degree of nuclear spin polarization of the up (down)-spin domain which are measured in terms of a change in the hyperfine field. The degree of spin polarization (resulting from the dynamical spin flip-flop process between the electron and nuclear spins driven by the source-drain current $I_{\text{sd}}$) in each domain increases with larger pumping time $\tau_{\rm{p}}$ reaching a maximum of $6.6\%$ ($4.0\%$) respectively. Without such pumping the thermal equilibrium polarization is much less than $1\%$.

Having described one half of our hybrid system, let us now describe the second system which is a NG boson. The NG boson arises from a charge imbalance in the QH state of the bilayers. We can tune this charge imbalance to realize three different spin phases / configurations: the canted antiferromagnetic (CAF) phase, the ferromagnetic (FM) phase and the spin-singlet (SS) phase (see supplementary material section II on how to access a particular spin configuration) \cite{Brey1999,Kumada2006}. It is the CAF phase that is of interest here as it supports a linear dispersing NG boson in the long wavelength limit \cite{Sarma1997}. The NG boson naturally couples with the  double-nuclear-spin polarized domains  \cite{yhamaetalDicke} allowing us to create our required  hybrid system (see supplementary material section on how to initialize and calibrate the nuclear spins). 

In Fig \ref{Fig02}b we show the effect of coupling our double domain system  to the NG boson for a given pumping time before measuring the total magnetization of the system (which indicate the degree of polarization of the hybrid system). It is clear that this coupling has rapidly increased the degree of polarization in the system as shown by the black dot curve - actually to a magnetization below that associated with thermal equilibrium. This leads to the natural question about what our system is doing and so let us now explore the dynamics of this system. We let the double nuclear spin system and NG boson interact for a given dwell time before we measure the polarization of the entire double domain system using a simple magnetization measurement. We clearly observe (as shown in Fig \ref{Fig02}c) that the nuclear spins from the double domain system have relaxed very rapidly (< $5$ ms) and are pointing upwards (parallel to the direction of the applied magnetic field). They stay in the relaxed state until around $1$s where it begins to thermalize to its original thermal steady state. The re-thermalization is completed near $100$ s. A sudden reconfiguration of nuclear spin polarization and its dynamics towards equilibrium are completely different from an ordinarily independent relaxation process.

It is apparent from Fig \ref{Fig02}c that our hybrid system has a number of interesting and independent timescales associated with it. Two of these are associated with the usual dephasing \cite{Paget1997,Sundfors1969} \& thermalization processes with timescales given by $T_{2^*}\sim 1$ms ($T_{1}\sim 40$s) respectively \cite{Ota2007}. These correspond clearly to the behaviour seen in the right hand side of that sub-figure. The left hand side of the figure, i.e. the flat region in between $5$ and $500$ ms, is however much more interesting and only occurs because of the coupling to the NG boson associated with the  CAF phase. The FM and SS phases do not show this short term behaviour. This is due to the fact that they have gapped modes \cite{yhamaetalDicke} meaning the double nuclear spins exhibit normal relaxation processes independent of the initial number of polarized nuclear spins \cite{FauziPRB} (see supplementary material). It is thus clear our observed behaviour here is associated with the canted antiferromagnetic (CAF) phase and its associated NG boson. It is critical to determine how this relaxation caused by the coupling to the NG boson varies with the size of the total nuclear spin ensemble. We have already established in Fig \ref{Fig02}a that the hyperfine field strength (proportional to the number of polarized nuclear spins in the ensemble) increases with increasing pumping time $\tau_{\rm{p}}$. Thus we can prepare different size nuclear spin polarized ensembles which we can then let interact with the NG boson for $5$ms before measuring the resulting polarization. In Fig.\ref{Fig02}b we plot the total spin polarization (measured by the hyperfine field) against the pumping time $\tau_{\rm{p}}$ after its interaction with the NG boson.  
%\begin{figure}[htb]
%\begin{center}     
%\centering
%\includegraphics[width=1.0\linewidth]{Fig3new.eps}
%\end{center}
%\caption{Plot of the total nuclear spin polarization (measured by the strength in the  hyperfine field) for varying pumping time $\tau_{\rm{p}}$ after a $5$ms interaction the NG boson mode. } 
%\label{Fig03} 
%\end{figure}
We clearly observe that the nuclear spin polarization of the total double domain system increasing as the pumping time $\tau_{\rm{p}}$ gets larger. It is clear from Fig. \ref{Fig02}b-c that this short time  behaviour is associated with the form of the NG boson coupling to the double nuclear spin domain. As such we need to explore the natural of the NG boson and its coupling in a little more detail. 

Let us discuss the QH state again briefly. When we tune the two-dimensional QH state such that the in-plane rotational symmetry of electron spin is spontaneously broken,  an associated linear dispersing NG boson emerges. This can be described by a continuous wave-number vector $\boldsymbol{k}=(k_x,k_y)$ with wavelength $\sim 0.1 $mm for a nuclear-spin frequency near $10$ MHz. This long wavelength is very important as the  nuclear-spin separation is approximately $\sim 0.5$ nm((much shorter than that of the NG boson)  meaning those nuclear spins can couple collectively to the NG boson. This is our first hint at a collective effect where the NG boson is acting like a  reservoir \cite{yhamaetalDicke}. It also gives us a natural way to model our overall hybrid system. For simplicity, we assume that all the spins here are identical (only single species to be taken into account) with spin-1/2. 

%\section{System modeling}\label{model}

Our hybrid quantum system composed of two nuclear spin ensembles and the NG boson can be effectively described by the Dicke model where the NG boson act as a reservoir \cite{yhamaetalDicke}. This model indicates the generation of collective phenomena of nuclear spins. In actual systems, however, there are individual dissipative effects which break the collective phenomena. Examples include  a dipole-dipole interaction between nuclear spins which induces a dephasing effect and an individual coupling between spin and other reservoir like phonon leading to a $T_1$-time relaxation process. Given the presence of collective effect involving a quantum / thermal reservoir as well as individual nuclear spin dephasing and thermalization, it is quite natural to model this system by a nuclear spin Born-Markov type master equation  \cite{Carmichaeltxb}  of the form 
%\begin{widetext}
%\begin{eqnarray}
%\dot \rho(t)&=& -i \omega_{\text{ns}} [ J^z_{\text{A}}+J^z_{\text{B}} ,   \rho(t) ]  + \frac{\gamma^{\text{rel}}}{2} \left[ (\bar{n}+1) {\cal L}(J^-_{\text{A}}+J^-_{\text{B}})+ \bar{n} {\cal L}(J^+_{\text{A}}+J^+_{\text{B}})\right]   \notag\\
%&+&  \frac{\gamma^{\text{rel}} }{2}    \Big{[} (\bar{n}+1)  \sum_{i_{\text{A}}=1} ^{N_{\text{A}}} {\cal L}(I^-_{i_{\text{A}}} + I^-_{i_{\text{B}}})  +\bar{n} \sum_{i_{\text{B}}=1} ^{N_{\text{B}}} {\cal L}(I^+_{i_{\text{A}}} + I^+_{i_{\text{B}}}) \Big{]}  +  \frac{ \gamma^{\text{dep}}  }{2}   \left[  \sum_{i_{\text{A}}=1} ^{N_{\text{A}}} {\cal L}(I^z_{i_{\text{A}}}) + \sum_{i_{\text{B}}=1} ^{N_{\text{B}}} {\cal L}(I^z_{i_{\text{B}}}) \right] ,
%\label{smasterequation1}
%\end{eqnarray}
%\end{widetext}
\begin{eqnarray}
\dot \rho(t)&=& -i \omega_{\text{ns}} [ J^z_{\text{A}}+J^z_{\text{B}} ,   \rho(t) ]  \notag\\
& + & \frac{\gamma^{\text{rel}}}{2} \left[ (\bar{n}+1) {\cal L}(\left[J^-_{\text{A}}+J^-_{\text{B}}\right]\rho)+ \bar{n} {\cal L}(\left[J^+_{\text{A}}+J^+_{\text{B}}\right]\rho)\right]   \notag\\
&+&  \frac{\gamma^{\text{rel}}}{2}    \Big{[} (\bar{n}+1)  \sum_{i_{\text{A,B}}=1} ^{N_{\text{A,B}}} {\cal L}(I^-_{i_{\text{A,B}}}\rho)  +\bar{n}\sum_{i_{\text{A,B}}=1} ^{N_{\text{A,B}}} {\cal L}(I^+_{i_{\text{A,B}}}\rho) \Big{]}  \notag\\
&+&  \frac{ \gamma^{\text{dep}} }{2}   \left[  \sum_{i_{\text{A}}=1} ^{N_{\text{A}}} {\cal L}(I^z_{i_{\text{A}}}\rho) + \sum_{i_{\text{B}}=1} ^{N_{\text{B}}} {\cal L}(I^z_{i_{\text{B}}} \rho) \right] ,
\label{smasterequation1}
\end{eqnarray}
where $J^z_{\text{A,B}} = \sum_{i_{\text{A,B}}=1} ^{N_{\text{A,B}}} I^z_{i_{\text{A,B}}} $ are the collective spin Z-operators for the domain A (B) with $I^z_{i_{\text{A,B}}}$ representing the individual nuclear spin 1/2 z-operator. Here $N_{\text{A,B}}$ are the total number of spins included in the domain A (B) respectively (the combined number of spins in both ensembles is $N=N_{\text{A}}+N_{\text{B}}$). Associated with these are the collective (individual) raising $J^+_{\text{A,B}}$ ($I^+_{i_{\text{A,B}}}$) and lowering  $J^-_{\text{A,B}}$ ($I^-_{i_{\text{A,B}}}$) operators for the nuclear spins in domain A, B respectively.The  nuclear spin frequency is given by $ \omega_{\text{ns}} = \gamma_{ \rm{n} } B$ where  $\gamma_{ \rm{n} }$  being the gyromagnetic ratio of nuclear spins. Further 
$\gamma^{\text{rel}}$ ($\gamma^{\text{dep}}$) in our master equation represent the damping (dephasing) rates. Next $\bar{n}=1/(e^{\hbar\omega_{\text{ns}} /k_{\text{B}}T}-1)$ is the Bose-Einstein distribution functions at the energy $\hbar\omega_{\text{ns}}$ for a given temperature $T$ where $k_{\text{B}}$ is the Boltzmann constant. In (\ref{smasterequation1}) the Liouvillian ${\cal L}(X\rho)$ is  given by ${\cal L}(X\rho)= 2X \rho X^\dagger - X^\dagger X \rho - \rho X^\dagger X$ with $X$ being an arbitrary operator. 

\begin{figure}[htb] 
\includegraphics[width=0.4\textwidth]{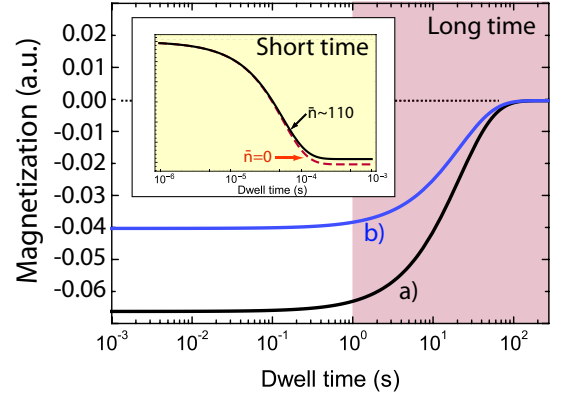}
\caption{Simulation of the total magnetization of the double domain nuclear spin versus the interaction (dwell) time, nothing that we have an arbitrary scaling on the y-axis. The results are presented for a) $N=10^6$, b) $N=6*10^5$. Here we have used the parameters $\omega_{\text{ns}}/ 2 \pi = 10 $ MHz, $T= 50$ mK,  $T_2^*\sim 1$ ms and $T_1 \sim 40$ s. This implies that ${\bar n} \sim 110$ which is much less than $N$.  Two distinct regions are shown in the main figure: white (red) where individual spin dephasing (thermalization) are the dominant behaviours. The yellow coloured inset shows the short time behaviour where collective thermalization effects dominate. Within the inset the black (red-dashed) curves correspond to (thermal mean photon number ${\bar n} \sim 110$ (zero temperature ${\bar n} \sim 0$) respectively. Little difference is seen for the two ${\bar n}$ as both are much less than $N$.} 
\label{simulation} 
\end{figure}

Our master equation given by (\ref{smasterequation1}) describes four basic phenomena; the Larmor precession of nuclear spins; collective thermalization, individual thermalization and individual dephasing (we ignore collective dephasing effects here at present). Now solving the master equation allows us to explore the dynamics of the overall system and we should be able to simply determine if we can reproduce the behaviour observed in Fig  \ref{Fig02}c - especially as we already know many of our system parameters including  $\omega_{\text{ns}}$ and the fridge temperature $T$ as well as the $T_1, T_2^*$ relaxation times, Our interest here will be in determining the magnetization of the double domain system which will be proportional (up to an arbitrary scaling) to the expectation value $ S_z= \langle J^z_{\text{A}} + J^z_{\text{B}} \rangle$.  In Fig. \ref{simulation}, we show the evolution of the total spin magnetization  versus dwell time for a small sized nuclear spin ensemble (not the exact number present in our actual system which will be approximately $10^{12}$). We do however to ensure that the total number of spins in our model is much greater than $\bar n$, otherwise the system dynamics may change. Here we choose $N=10^6$. With our chosen parameters it is clear that we have a number of distinct behaviour arising over different time scale. Those correspond to time scale less than $1$ ms (yellow shaded region), between $1 -1000$ ms and greater than $1$ s (red shaded region).   The main figure in Fig. \ref{simulation} is consistent with the experimental observations of Fig. \ref{Fig02}(c).
We need to explore these three regions in a little more detail starting with the short time regime shown in the inset for a total spin number $N=10^6$. First and foremost the decay process we observe here occurs much faster than either the $T_2^*$ or $T_1$ times associated with individual nuclear spins. The collective thermalization term in our master equation does reproduce such behaviour giving us strong evidence of collective and coherent effects here. Next the individual dephasing of the nuclear spins with associated $T_2^*\sim 1$ ms dominates in the $1 -1000$ ms temporal region where it destroys all coherence within and between the nuclear spins leaving the double domain system in a separable state. Each of the nuclear spins are now acting independently of each other and so collective effects do not arise. At approximately $1$ s  the effect of individual nuclear spin thermalization begins and at approximately $100$ s the nuclear spins reach their high temperature steady state (with a magnetization of zero).  A similar behaviour (blue  curve in Fig. \ref{simulation}) occurs if $N$ decreases but with a different degree of short time total magnetization is reached. The long time steady state is the same. Our model naturally explains the observed experimental behaviour and can give us further incites into our system dynamics.

The particularly interesting temporal behaviour arises in the short time regime where we observe a fast decrease in the total magnetization to a little well below zero (the magnetization associated with the long time steady state). We only observe the resultant effect here and not it dynamics which we can not experimentally measure. We only see the result that this collective behaviour has caused. Our model however can also be used to explore the short time regime in more theoretical detail. It is clear from our model (with ${\bar n}=0$ that the rate of thermalization scales quadratically with $N$ (rather than linear from individual spin thermalization). This is consistent with superradiance decay. Next for ${\bar n} \sim 110$ we observe a very similar quadratic short time behaviour, however as $ {\bar n} \rightarrow N$ we loss that quadratic behaviour. In our experiment we realistically have $N>10^{12}$ which is much greater than  the ${\bar n} \sim 110$ associated with the nuclear spin operating at $50$ mK. For such a large $N$ the associated superradiant decay time is extremely fast ($<10^{-10}$ s) and hence is impossible to observe. Our observed experimental behaviour however is only consistent if such collective decay had occurred.

%The inset shows the short time where we clearly observe the magnetization's rapid decay (on a time scale much much faster than $T_1$. In the region [1 - 1000 ms] single spin dephasing destroys any coherence between spins meaning any collective effects including entanglement have disappeared. From 1 s onwards we observe the normal thermalization process associated with the individual nuclear spins. 
%present the time evolution relaxation processes of normalized total collective double nuclear spin domains defined by $ S_z /N = \langle J^z_{\text{A}} + J^z_{\text{B}} \rangle / ( N_{\text{A}} + N_{\text{B}} )$ for various $N_{\text{A,B}}\sim 10^6$ 

%\red{ with $  \langle X \rangle = \text{Tr} (X \rho(t) ). $ It is obtained by solving Eq. \eqref{smasterequation1} numerically.  Here we take $ N_{\text{A}} =25, N_{\text{B}} = 15, \gamma^{\text{rel}}_{\text{col}} = X, \gamma^{\text{dep}} _{\text{ind}} = XX, \gamma^{\text{rel}}_{\text{ind}} = XXX, \bar{n}_{\text{col}} = x,$ and  $ \bar{n}_{\text{ind}} = xx.$}

%To show the dynamics structure of double-nuclear-spin domains clearly we split the $ s^z $ curve into two time regimes.   The blue curve represents the short-time-scale dynamics which ends at $t = 0.5$ sec and the red curve is the long-time-scale dynamics of nuclear spins. Let us discuss the characteristics of nuclear spin dynamics.

%We ignore collective dephasing for the moment as it would by itself not explain the short time behaviour observed.

To summarize, we have shown how collective and coherence effects arise in a double domain nuclear spin ensemble coupled to the Nambu-Goldstone boson in a GaAs semiconductor. The NG bosons long wavelength means both nuclear spin ensembles couple see the same mode. This leads to superradiant like decay event when our system is operating in the reigme ($\hbar\omega_{\text{ns}} \ll k_{\text{B}}T$). Further it is likely that the short time dynamics will show entanglement between the two ensembles. This will open  a new paradigm in nonlinear systems where quantum only (rather than coherent) effects are present.

%The blue curve represents the dynamics governed by the collective decay channel. It generates the steady (thermal-equilibrium)  state  where the value of $ s^z$  depends on both $ N_{\text{A}} $ and $ N_{\text{B}}$.The plateau which starts from $t = 0.01$ sec and reaching  $ t= 0.1$ sec corresponds to this first steady state.   Around $t = 0.5$ sec, which is the starting point of red curve,  the individual dephasing effect starts to compete with the collective-decay effect.   Then at $ t= 0.1$ sec, the collective behavior of double-nuclear-spin-domains becomes broken. The dynamics is going to be governed by the individual dephasing and the individual decay channels. From $ t \sim 60$ sec, what we see is the formation of second plateau. It describes the generation of second steady state where  $ s^z$ does not depend on $ N_{\text{A}} $ and $ N_{\text{B}}$ but on  $\bar{n}_{\text{ind}}$ (or the temperature $T_{\text{ind}}$).  As a result, when there are both collective and individual dissipative channels the nuclear spins show double relaxations associated with the two different plateaus (steady states) are created;     the first plateau is induced by the collective decay while the second one by the individual dephasing and relaxation effects.  By comparing the relaxation curves in Fig. \ref{simulation} and those in Figs. \ref{Fig02}(b) and \ref{Fig03}, we consider that our modeling  explains qualitatively the characteristics as well as the mechanism of nuclear spin relaxation processes of our experimental data.  

\acknowledgements
We thank Yusuke Hama and Shane Dooley for valuable discussions. This work was supported in part by the MEXT Grant-in-Aid for Scientific Research on Innovative Areas KAKENHI Grant Number JP15H05870 (Y.~H, E.~Y, and K.~N), MEXT Grant-in-Aid for Scientific Research(S) KAKENHI Grant Number JP25220601 (E.~Y and K.~N) and Grant-in-Aid for Scientific Research(A) KAKENHI Grant Number JP19H00662 (W.J.M and K.~N).

%\section*{author contributions}
%Y.H conceived the idea and supervised the project. M.H.F fabricated the devices and carried out the measurements. W.J.M and K.N performed numerical simulations. All authors discussed, analyzed the experimental data, and wrote the manuscript. 

%\section*{Competing interests}
%The authors declare no competing interests.


\begin{thebibliography}{99}
\bibitem{footnote1}
We observe a slight asymmetry between the numbers of up and down nuclear spin polarization after current pumping. The nature of this asymmetry  is still an open question, however,  it is likely to stem from the spatial topology of electron spin domains used to dynamically polarize the nuclear spins \cite{Nick}.
\end{thebibliography}
\end{document}

%\preprint{APS/123-QED}

\title{Supplementary Materials for Double Nuclear-Spin Relaxations in Hybrid Quantum Hall Systems}% Force line breaks with \\

\author{M.~H.~Fauzi}
\affiliation{Center for Spintronics Research Network, Tohoku University, Sendai 980-8577, Japan}
\affiliation{Research Center for Physics, Indonesian Institute of Sciences, South Tangerang City, Banten 15314, Indonesia}
%\email{moha065@lipi.go.id}

%\author{Y.~Hama}
%\affiliation{National Institute of Informatics, 2-1-2 Hitotsubashi, Chiyoda-ku, Tokyo 101-8430, Japan}

\author{W. J. Munro}
\affiliation{National Institute of Informatics, 2-1-2 Hitotsubashi, Chiyoda-ku, Tokyo 101-8430, Japan}
\affiliation{NTT Basic Research Laboratories \& NTT Research Center for Theoretical Quantum Physics, NTT Corporation, 3-1 Morinosato-Wakamiya, Atsugi-shi, Kanagawa, 243-0198, Japan}

\author{K. Nemoto}
\affiliation{National Institute of Informatics, 2-1-2 Hitotsubashi, Chiyoda-ku, Tokyo 101-8430, Japan}

\author{Y. Hirayama}
\affiliation{Center for Spintronics Research Network, Tohoku University, Sendai 980-8577, Japan}
\affiliation{Department of Physics, Tohoku University, Sendai 980-8578, Japan}
\affiliation{Center for Science and Innovation in Spintronics (Core Research Cluster), Tohoku University, Sendai 980-8577, Japan}
%\email{yoshiro.hirayama.d6@tohoku.ac.jp}

\date{\today}% It is always \today, today,

%\begin{abstract}
%An article usually includes an abstract, a concise summary of the work
%covered at length in the main body of the article. 
%\begin{description}
%use the optional argument of the \verb+\item+ command to give the %category of each item. 
%\end{description}
%\end{abstract}

\maketitle

%\tableofcontents

\section{Device and measurement details}

\begin{figure}[t]
\begin{center}    
\centering
\includegraphics[width=0.7\linewidth]{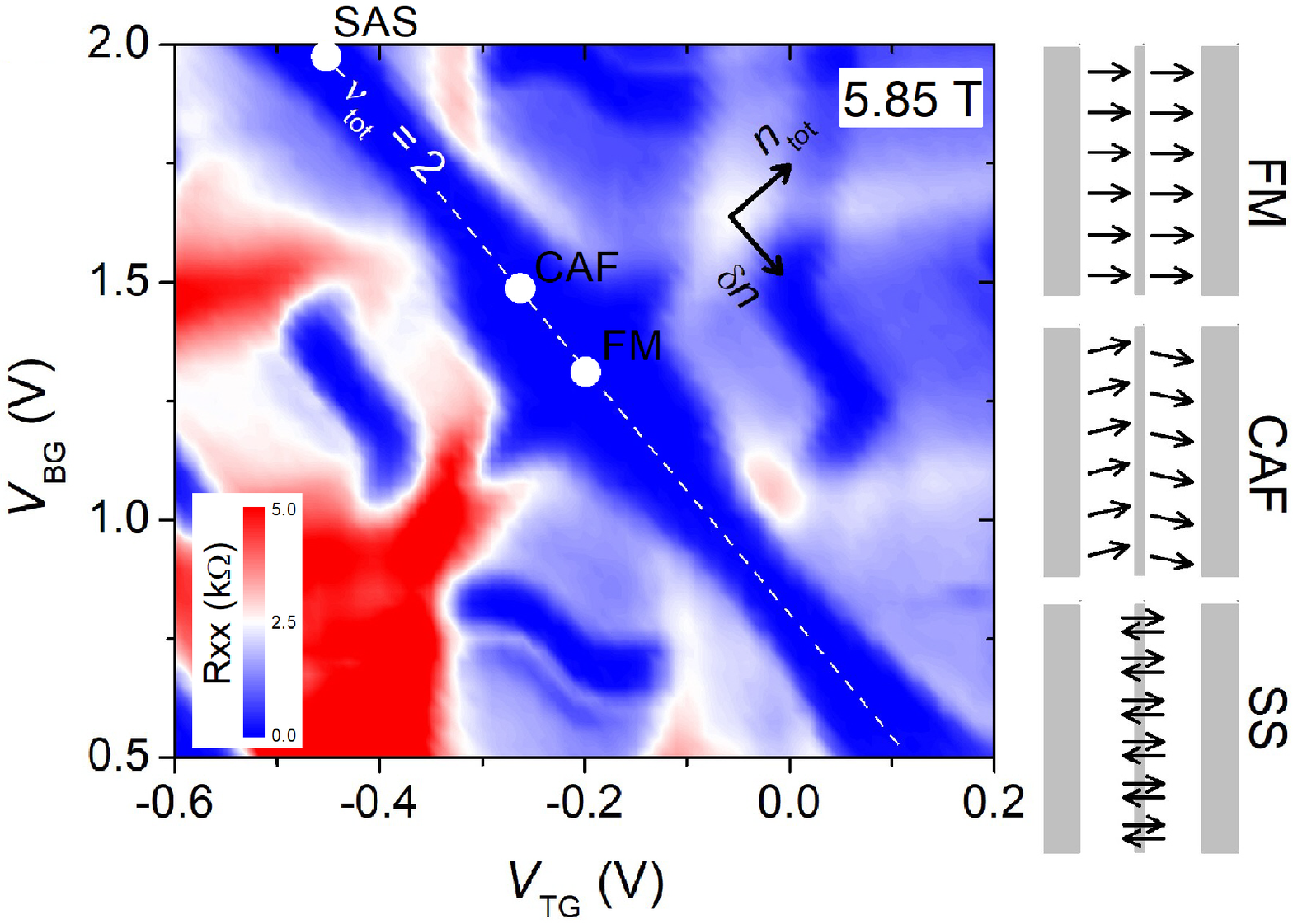}
\end{center}
\caption{A color map of longitudinal resistance as a function of top and bottom gate measured at a field of $5.85$ T and a lattice temperature of $50$ mK. We highlight the data around total filling factor $\nu_{\rm{tot}} = 2$ where three different electron spin configurations (FM, CAF, and SS) along the dotted white line can be realized depending on the charge imbalance between the two layers.}
\label{FigS1} 
\end{figure}

MBE grown two identical $20$-nm wide GaAs quantum wells are brought in a close proximity separated by a $2.2$-nm thick insulating Al$_{0.33}$Ga$_{0.67}$As spacer, resulting in a rather strong tunneling gap of about $15$ K. The wafer is photo-lithographically carved into a $30$ $\mu$m-wide Hall bar geometry as schematically shown in the main text. The electron density as well as the charge imbalance between the two layers can be controlled by gating through the Ti/Au metal top gate and bottom gate. The samples were measured in a dilution refrigerator with a lattice temperature of $50$ mK. All measurements were carried out using a standard lock-in technique at $333$ Hz. We set the signal acquisition time long enough to account for the lock-in integration time.

\section{Spin configuration at $\nu_{\rm{tot}} = 2$}

Depending on the charge imbalance between the two layers, three different spin configurations shown at the right hand side of the Fig. S\ref{FigS1} are expected to realize in bilayer quantum Hall at a total filling factor $\nu_{\rm{tot}} = 2$, namely ferromagnetic (FM), canted antiferromagnetic (CAF), and spin singlet states (SS). The spin configuration can be tuned by applying charge imbalance between the two layers, controlled by the top and bottom gates. In the absence of charge imbalance, the electron spin in the two layers possesses ferromagnetic configuration (FM). In contrast, for a large charge imbalance, the spin configuration changes to spin single state (SS). In between the two cases emerges a new spin state where the spin in each layer is canted with respect to the magnetic field (CAF). While the FM and SS states are gapped, the canted state is predicted to support a linearly dispersing Goldstone mode whose excitation energy is gapless in the long wavelength limit. We use this particular spin state as a bosonic reservoir for a double nuclear spin domains. To set the electronic state to the canted antiferromagnetic state we tune the top gate bias $V_{\rm{TG}} = -0.23$ V and bottom gate bias $V_{\rm{BG}} = +1.5$ V.
%\nocite{*}

\section{Initialization}

\begin{figure}[t]
\begin{center}    
\centering
\includegraphics[width=0.9\linewidth]{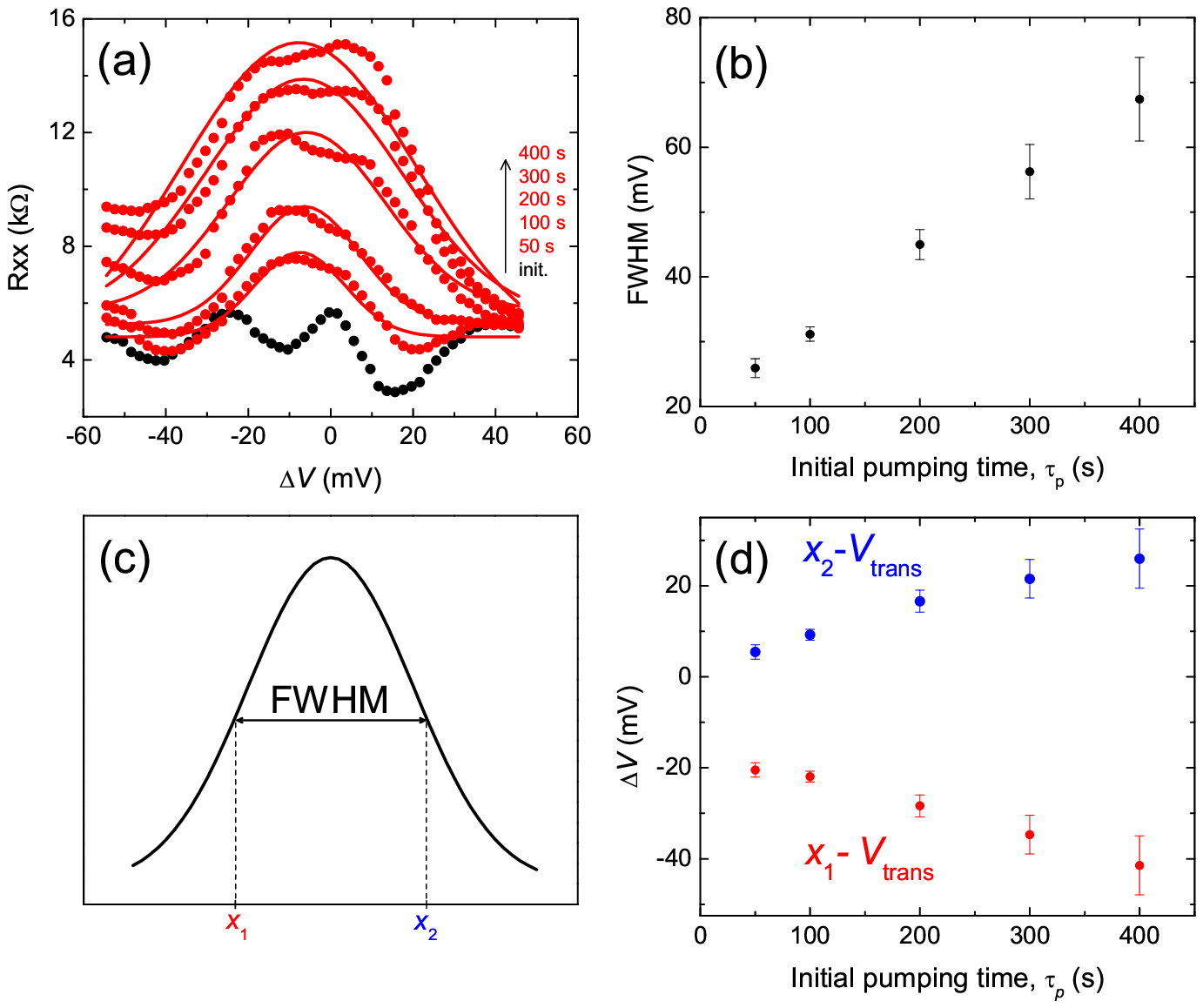}
\end{center}
\caption{(a) Evolution of longitudinal resistance at the $\nu=2/3$ spin transition recorded after current-induced nuclear spin polarization from $50$ to $400$ seconds at the top gate bias $V_{\rm{trans}} = -1.326$ V with $\Delta V \equiv V_{\rm{TG}} - V_{\rm{trans}}$. The black dot curve is the spin transition recorded before current-induced nuclear spin polarization. The red line is a Gaussian fit to the data. (b) Full-width at half maximum (FWHM) as a function of initial pumping time extracted from panel (a). (c) Schematic illustration of how to extract the relocated spin transition (\textcolor{red}{$x_1$} and \textcolor{blue}{$x_2$}) from FWHM. (d) The relocated spin transition with respect to $V_{\rm{trans}}$ as a function of initial polarization time extracted from the FWHM in panel (b).}
\label{FigS2} 
\end{figure}

We use the spin transition at the fractional filling $\nu=2/3$ to initialize the double nuclear spin domain as shown in Fig. S\ref{FigS2}(a). It is important to note that the nuclear spin domains with different spin orientation (pointing up and down) are initialized and polarized only at the bottom layer. The top layer is completely depleted during the initialization and readout. To accomplish this, we tune the bottom layer electronic state to the $\nu=2/3$ spin transition point by setting the bottom gate bias to $V_{\rm{BG}} = +1.5$ V and the top gate bias to $V_{\rm{trans}} = -1.326$ V. This bias setting corresponds to $\Delta V = 0$ point in Fig. S\ref{FigS2}(a). We then apply a source-drain current of $60$ nA with a varying keeping time from $50$ to $400$ seconds as indicated in the label of Fig. S\ref{FigS2}(a) at $\Delta V = 0$ point. For the readout, we record the longitudinal resistance evolution after current-pumping by scanning the top gate $V_{\rm{TG}}$ around the spin transition point while the bottom gate bias is kept constant at $V_{\rm{BG}} = +1.5$ V. In other word, the $\nu=2/3$ spin transition peak at the bottom layer is measured as a function of the top gate (represented by $\Delta V$) instead of as a function of the bottom gate, as shown in Fig. S\ref{FigS2}(a). The resistance around the spin transition gets bigger and wider with increasing the current-pumping time, indicating dynamic nuclear polarization build up with increasing the pumping time $\tau_p$.

To extract the nuclear polarization for different keeping time, we first evaluate the full-width at half-maximum (FWHM) of each resistance evolution around the transition point as plotted in Fig. S\ref{FigS2}(b). The increase reflects an increase in the up and down nuclear spin polarization in the bottom layer. The up nuclear spin polarization relocates the spin transition to the left ($\Delta V < 0$) and likewise the down nuclear spin polarization relocates the spin transition to the right ($\Delta V > 0$). 

After evaluating the FWHM, we then extract the relocated spin transition after current-induced dynamic nuclear polarization. To do so, we employ the procedure schematically illustrated in Fig. S\ref{FigS2}(c). The corresponding relocated spin transition with respect to the initial spin transition ($V_{\rm{trans}}$) for each data subset are displayed in Fig. S\ref{FigS2}(d). The red and blue points correspond to the relocated spin transition due to up and down nuclear spin polarization, respectively.

This relocated spin transition, quantified as a shift in the top gate voltage, can be directly converted to the hyperfine field as we discuss in the next section.

\section{Gate voltage conversion to Hyperfine Field}

\begin{figure}[t]
\begin{center}    
\centering
\includegraphics[width=0.5\linewidth]{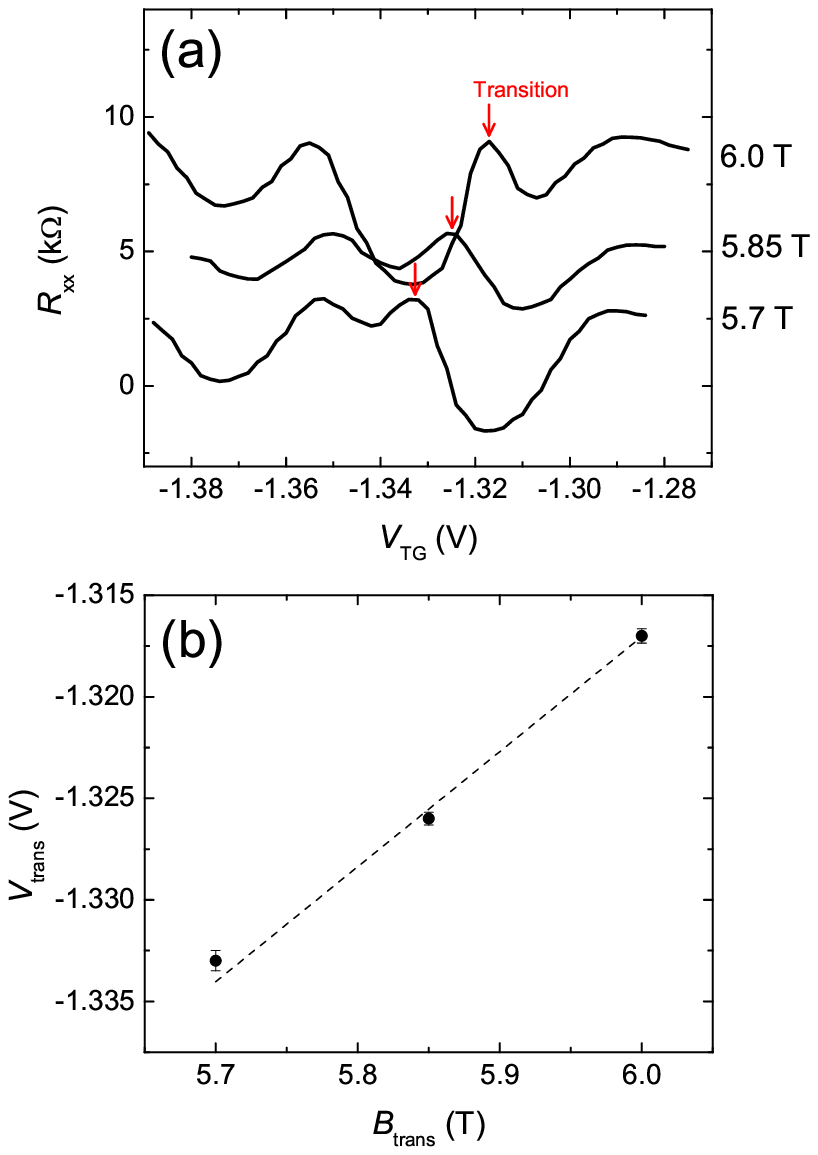}
\end{center}
\caption{(a) The $\nu=2/3$ spin transition point recorded at three different magnetic fields. The transition is seen as a protrusion in the longitudinal resistance marked by the red arrow. Note that the transition is recorded in the absence of current-pumped nuclear spin polarization. (b) The transition point extracted from panel (a) follows a linear relation.}
\label{FigS3} 
\end{figure}

At a fix magnetic field, in response to the hyperfine field coming from nuclear spin polarization, the spin transition will get relocated ($\Delta V \neq 0$) as discussed earlier. To be able to convert the shift in the gate voltage to the hyperfine field, we track the spin transition position at three different magnetic fields in the absence of current-pumping as displayed in Fig. S\ref{FigS3}(a). The transition point, marked by the red arrow, is relocated to a less negative top gate bias voltage with increasing the field. The corresponding transition point for each data set in Fig. S\ref{FigS3}(a) is displayed in panel (b) of Fig. S\ref{FigS3}, from which a linear relation is obtained ($V_{\rm{trans}} = 0.0533B_{\rm{trans}} -1.638 $). 
If the relocated spin transition to a different gate voltage at $5.85$ T due to nuclear spin polarization coincides with the spin transition at a particular magnetic field $B_{\rm{trans}}$ as displayed in Fig. S\ref{FigS3}(b), then the amount of hyperfine field can be estimated from the following relation \cite{AkibaAPL}
\begin{equation}
    B_N (\rm{T}) = \sqrt{5.85B_{\rm{trans}}} - 5.85.
\end{equation}
Substituting the linear relation between $B_{\rm{trans}}$ and $V_{\rm{trans}}$ extracted from Fig. S\ref{FigS3}(a) into equation (1), we get
\begin{equation}
    B_N (\rm{T}) = \sqrt{109.756(V_{\rm{trans}} + 1.638)} - 5.85.
\end{equation}
This final equation is used to convert the relocated spin transition into the hyperfine field. Note that $B_N = 0$ T at $V_{\rm{trans}} \approx -1.326$ V. If the spin transition relocates to a less negative bias voltage ($V_{\rm{trans}} < -1.326$ V) then the hyperfine field has a negative value and vice versa.
%Note that when $B_{\rm{trans}} = 5.85$ T (the spin transition does not relocate), the equation returns $B_N = 0$ as expected.

\section{Nuclear Spin Dynamics due to interaction with a gapless and gapped electronic state: A comparison}

\begin{figure}[t]
\begin{center}    
\centering
\includegraphics[width=\linewidth]{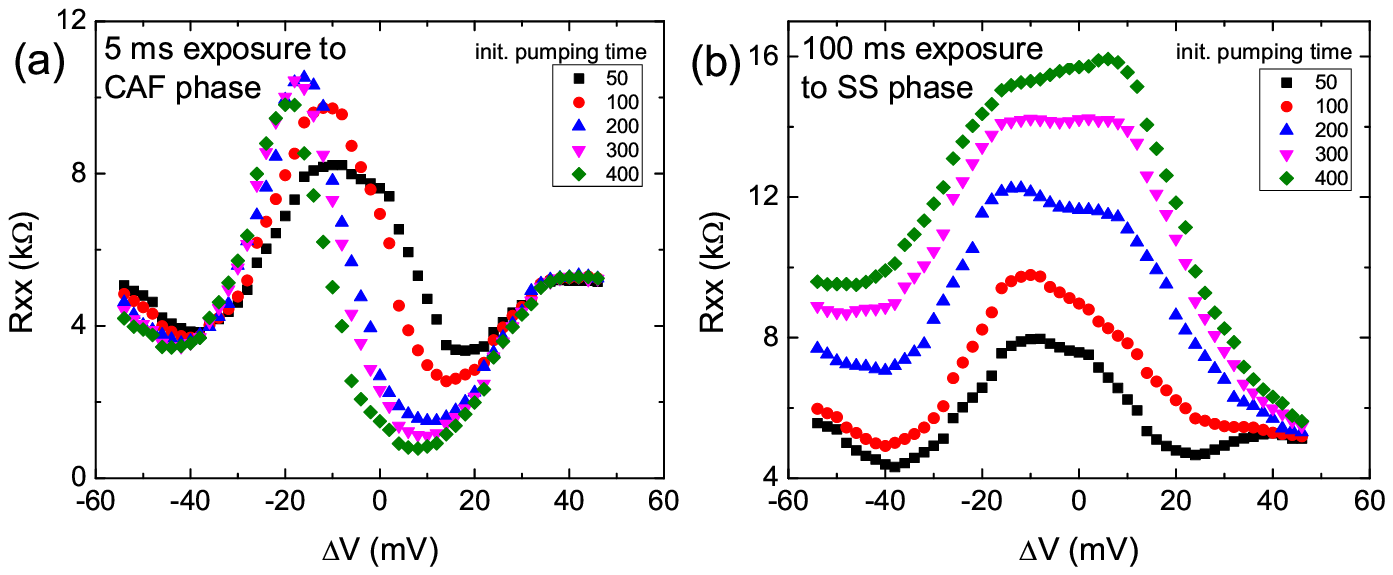}
\end{center}
\caption{Longitudinal resistance $R_{\rm{xx}}$ swept around the $\nu=2/3$ spin transition recorded after (a) $5$ msec exposure to canted antiferromagnetic and (b) $100$ ms exposure to spin-singlet phases for five different initial pumping times.}
\label{FigS4} 
\end{figure}

For the sake of clarity, here we show raw data of nuclear spin polarization dynamics due to interaction with CAF and SS phases for a $5$ and $100$ ms exposure, respectively. One can see differing responses already just by looking at the raw data displayed in Fig. S\ref{FigS4}. The brief exposure to CAF phase dramatically change the recorder spin transition as displayed in Fig. S\ref{FigS4}(a). The transition curve now has a very narrow width than the initial spin transition after current-pumping with its transition point relocates to the left hand side ($\Delta V < 0$). The longer the current-pumping time, the narrower the transition curve becomes.

In contrast, we do not see any dramatic effect when the nuclear spin polarization interacts with a gapped SS phase for a $100$ ms exposure. Note that the vast majority of electronic states in the quantum Hall regime are gapped, so that the nuclear spin relaxation dynamics would follow an ordinary classical relaxation process like the one shown in Fig. S\ref{FigS4}(b).

%a $5$ msec exposure to the canter antiferromagnetic phase displayed in Fig. 2(b) of the main text. Comparing the spin transition recorded after current-pumping discussed in section III, we see a dramatic change to the recorded spin transition following the brief exposure as shown in Fig. S\ref{FigS4}. The transition curve now has a very narrow width than the initial spin transition after current-pumping with its transition point relocates to the left hand side ($\Delta V < 0$). The longer the current-pumping time, the narrower the transition curve becomes.

%What does the figure tell us?. It tells us that the remaining nuclear polarization after the brief exposure to the canted antiferromagnetic phase has changed from inhomogeneous with a double nuclear spin domain to homogeneous nuclear spin polarization with a single nuclear spin domain, with the nuclear spin now pointing parallel to the magnetic field.

%\section{Nuclear Spin Dynamics due to interaction with a gapped electronic state}

%The classical and collective effect are distinct from one another and can be distinguished in our experiment as displayed in the following figures. The one on the left is the recorded behaviour after interaction with CAF phase for 5 ms for five different initial pumping time, as displayed in the supplementary Fig. S4. The one on the right is the behaviour after interaction with gapped SS phase for 100 ms.

\begin{figure}[t]
\begin{center}    
\centering
\includegraphics[width=\linewidth]{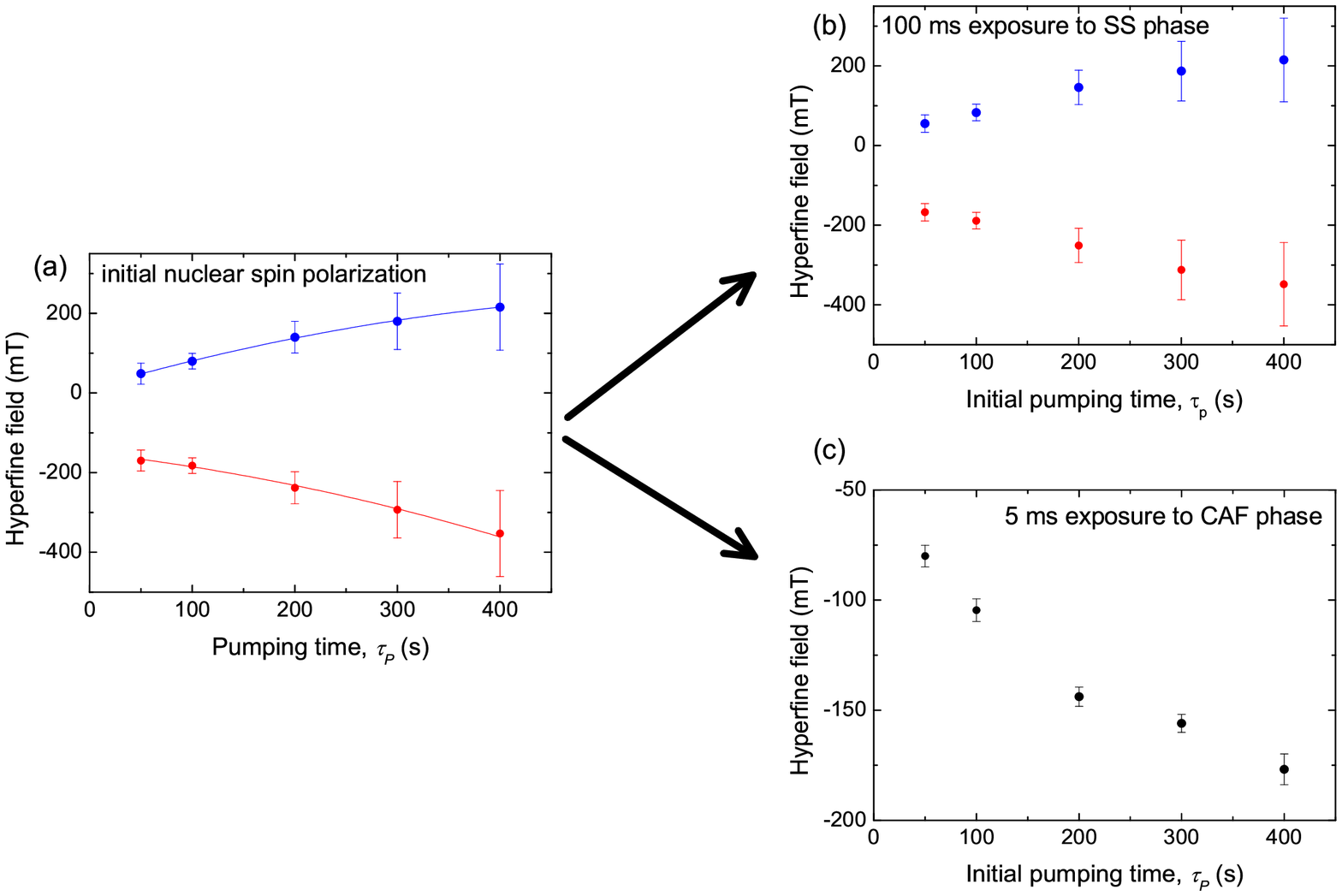}
\end{center}
\caption{Longitudinal resistance $R_{\rm{xx}}$ swept around the $\nu=2/3$ spin transition recorded after (a) $5$ ms exposure to canted antiferromagnetic and (b) $100$ ms exposure to spin-singlet phases for five different initial pumping times.}
\label{FigS5} 
\end{figure}

Now plotting the raw data into the hyperfine field vs pumping time as we did in the main manuscript displayed in Fig. S\ref{FigS5}, the two contrasting dynamics become much clearer. The nuclear spin configuration still very much resembles the initial configuration after $100$ ms interaction with the SS phase for all initial pumping times. Both nuclear spin domains relax towards equilibrium very slowly. In contrast, displayed in Fig. S\ref{FigS5}(c), the interaction with CAF phase for $5$ ms reconfigures the nuclear spin polarization dramatically from its initial state and shows pumping time dependence not seen in the ordinary case. 

%Although a rapid nuclear spin relaxation rate has been observed reported in Science [33] and the subsequent report in PRL 2007 (PRL 99 076805). However the way that the nuclear spin dynamics was measured missed out a key feature, which we address here namely a sudden reconfiguration of nuclear spin polarization after a brief interaction with CAF phase.

%Here, $B_{\rm{trans(unpol)}} = 5.85$ T is the field at which $B_{N} = 0$. $B_{\rm{trans(pol)}}$ corresponds to the field at which the transition point shifts in the gate voltage $V_{\rm{TG}}$ at $B_{\rm{trans(unpol)}} = 5.85$ T and $B_{N} \neq 0$.

%\begin{equation}
%    B_N = \sqrt{B_{\rm{trans(unpol)}}B_{\rm{trans(pol)}}} - B_{\rm{trans(unpol)}}
%\end{equation}
